\documentclass[amsmath,amssymb,prb,twocolumn,superscriptaddress,longbibliography,hypertext]{revtex4-2}

\usepackage{float}
\usepackage{siunitx}
\usepackage{booktabs}
\usepackage{graphicx}
\usepackage{array}
\usepackage{amsthm}
\usepackage{enumerate}
\usepackage[table]{xcolor}
\usepackage{bbm}
\usepackage{color}
\usepackage[normalem]{ulem}
\usepackage{multirow}
\RequirePackage[colorlinks=true,linkcolor=blue]{hyperref}
\usepackage{braket}

\begin{document}

\title{Nonreciprocal Acoustic and Optical Phonon Dispersion Mediated by Berry Curvature in Chiral Weyl Semimetals }
\author{Sanghita Sengupta}

\affiliation{Department of Chemistry, Brandeis University, Waltham, MA 02453}

\date{\today}                                       
\begin{abstract}
We investigate the phonon magnetochiral effect (PMCE) in chiral Weyl semimetals by deriving the nonreciprocal dispersion relations of both acoustic and non-polar optical phonons in the presence of a magnetic field. Using a semiclassical Boltzmann kinetic framework that incorporates Berry curvature, orbital magnetic moment, and node-dependent electronic structure, we obtain analytic expressions for the magnetic-field-induced corrections to the phonon dynamical matrix. Inequivalent Weyl nodes with distinct Fermi velocities, Fermi energies, and relaxation times generate a dynamical chiral imbalance that alters the phonon dispersion. For acoustic phonons, the formalism yields the magnetic-field-dependent corrections to the longitudinal mode, while for optical phonons we identify an optical analogue of the PMCE that produces a corresponding shift in the optical branch. Together, these results provide a unified theoretical description of how band-geometric properties of Weyl fermions influence both acoustic and optical phonon dispersions in chiral Weyl semimetals.
\end{abstract}

\maketitle

\section{Introduction}
\label{sec:intro}

Nonreciprocal responses characterize directional transport and propagation of quantum particles and quasiparticles such as electrons, phonons, photons, and spins. Such responses typically arise in materials that lack inversion symmetry and, in many cases, also break time-reversal symmetry through an external magnetic field or intrinsic magnetization. Broadly, nonreciprocal phenomena fall into two categories: linear and nonlinear effects. In the linear regime, prominent examples include the magnetochiral effect, nonreciprocal magnon transport, and chiral spin-current generation in systems that lack inversion symmetry and additionally break time-reversal symmetry through magnetization or an applied magnetic field \cite{Tokura2018, Ideue2017, takashima}. In the nonlinear regime, characteristic manifestations include unidirectional magnetoresistance in polar and chiral semiconductors, nonlinear Hall and shift-current responses, and photocurrents of topological origin \cite{Tokura2018, Ma2019, Sodemann2015}. Microscopically, many of these nonreciprocal behaviors originate from geometric quantities of the electronic structure such as the Berry curvature, toroidal moment, and magnetoelectric multipoles \cite{Tokura2018}.

Quantum materials exhibiting nonreciprocal responses are generally characterized by the absence of inversion and mirror symmetries, and often by broken time-reversal symmetry. When time-reversal symmetry remains intact, nonreciprocity can still arise through band-structure geometry, leading to linear effects such as natural optical activity and circular dichroism, as well as nonlinear phenomena including the shift current, the nonlinear Hall effect, and photovoltaic responses in non-centrosymmetric crystals \cite{Tokura2018,Sipe2000, Young2012, Morimoto2016, Sodemann2015}. These effects occur without the need for a magnetic field or magnetization. In contrast, when time-reversal symmetry is broken either by an external magnetic field or intrinsic magnetization, additional linear nonreciprocal effects emerge, such as the magnetochiral effect, the optical magnetoelectric effect, and nonreciprocal magnon transport \cite{Tokura2018, takashima}. In the nonlinear regime, broken time-reversal symmetry gives rise to responses such as nonlinear optical rectification, electrical magnetochiral anisotropy, the inverse Edelstein effect, and current-direction–dependent magnetoresistance \cite{Rikken2001, Tokura2018, Edelstein1990}.

One notable example of a nonreciprocal phenomenon is the phonon magnetochiral effect (PMCE), in which sound waves propagating parallel and antiparallel to an external magnetic field acquire different frequencies and attenuations. While the broader magnetochiral effect is ubiquitous occurring for electrons, photons, magnons, and other quasiparticles, the phonon magnetochiral effect is particularly difficult to observe because phonons couple only weakly to magnetic fields. PMCE was first experimentally identified in the chiral insulating ferrimagnet Cu$_2$OSeO$_3$, where the nonreciprocity was attributed to hybridization between acoustic phonons and chiral magnons \cite{Nomura2019}. More recently, PMCE has also been observed in a metallic chiral magnet, Co$_9$Zn$_9$Mn$_2$, demonstrating that phonon nonreciprocity can arise in itinerant systems as well \cite{Nomura2023}. These developments motivate the search for alternative mechanisms that go beyond magnon–phonon coupling and can enhance phonon nonreciprocity in conducting and topological materials.

In another class of systems, namely chiral Weyl semimetals \cite{huang2016, chang2017, tang2017, chang2018, gooth2019, rao2019, sanchez2019, schroter2019, takane2019}, a phonon magnetochiral effect has been theoretically predicted, demonstrating that nontrivial electronic band topology can imprint itself directly onto lattice dynamics. In this context, the Berry curvature and orbital magnetic moment of Weyl fermions modify the directional propagation of longitudinal acoustic phonons, giving rise to a PMCE even in the absence of magnon–phonon hybridization \cite{senguptapmce}. The prior study showed that longitudinal phonons propagating collinearly with an external magnetic field generate a dynamical chiral population imbalance, which in turn enhances the magnitude of the effect. A key outcome of this work was the identification of phonons as non-electronic probes of topological band invariants, extending beyond conventional electronic transport and photoemission techniques. These results motivate the exploration of additional phonon-based signatures of topological band geometry, including nonreciprocity in optical phonons and dispersion-level effects, which we pursue in the present work.

In this work, we extend the formalism developed in Ref.~[\onlinecite{senguptapmce}] to incorporate the effects of inequivalent Weyl nodes on phonon dispersion in the presence of a magnetic field, treating both acoustic and non-polar optical branches on equal footing. Recent studies have primarily focused on sound attenuation and acoustic magnetochiral dichroism \cite{antebi, sukhachov}, where nonreciprocal responses originate from differences in Fermi energies and Fermi velocities at distinct Weyl nodes. Here, we build upon these insights by analyzing how such node-dependent electronic properties modify the full phonon dispersion, and by extending the framework to include optical phonons. This reveals subtle but important variations that arise once the electronic structure of each Weyl node is treated individually. 

Our results show that phonons both acoustic and optical are sensitive non-electronic probes of topological band geometry, reflecting how momentum-space structure such as Berry curvature and orbital magnetic moment imprints itself on lattice dynamics. In particular, we show how differing Fermi velocities and Fermi energies at the Weyl nodes influence the dispersion relations of both acoustic and optical phonons, thereby giving rise to the phonon magnetochiral effect.

We employ a semiclassical framework that combines the Boltzmann kinetic equation for Weyl electrons with Maxwell’s equations for the electromagnetic fields induced by lattice motion and the elasticity equations governing phonon dynamics. Within this approach, we derive explicit expressions for the phonon dispersion, including both its real and imaginary components, that incorporate the effects of Berry curvature and the orbital magnetic moment of the electronic states. Although the formalism is broadly applicable, we focus here on chiral Weyl semimetals, where node-dependent band geometry plays a central role. The technical development of the method is presented in Sec.~\ref{sec:method}.

Beyond the band-geometric contributions, our analysis also highlights the role of a dynamical anomaly, a phonon-driven analogue of the chiral anomaly known from Weyl semimetals. In electronic transport, the chiral anomaly reflects the nonconservation of chiral charge in the presence of parallel electric and magnetic fields \cite{NielsenNinomiya1983,SonSpivak2013,Burkov2015,ArmitageRMP2018}. In the present context, longitudinal phonons propagating along a magnetic field generate effective, time-dependent electromagnetic fields that couple asymmetrically to opposite Weyl nodes, producing a dynamical chiral population imbalance. This non-equilibrium analogue of the anomaly feeds back into the phonon dynamics and enhances the magnetochiral response. The resulting lattice signatures provide a complementary probe of anomaly-related physics, which has traditionally been accessed through electronic transport and optical measurements. Thus, our work highlights phonons as sensitive detectors of the topological nature of chiral Weyl materials.

The rest of the paper is organized as follows. In Sec.~\ref{sec:method}, we introduce the physical ingredients of the model and outline the methodology, including the semiclassical formalism based on the Boltzmann kinetic equation and the elasticity equations that lead to the phonon dispersion relations. Section~\ref{sec:zero-mag} presents the influence of chiral electrons on phonons in the absence of a magnetic field, providing detailed derivations of the solution to the Boltzmann equation, the resulting drag force exerted by phonons on electrons, and its connection to the dispersion relations of both acoustic and optical modes. This section therefore establishes the zero-field electron–phonon renormalization of the phonon dispersion. In Sec.~\ref{sec:pmce}, we derive the dispersion relations in the presence of a magnetic field and examine the magnetochiral effects on both acoustic and optical phonons, including analytical and numerical results illustrating how the dispersion responds to changes in the direction of the magnetic field. Finally, Sec.~\ref{sec:discussion} summarizes our findings and discusses the origin of the effect as well as possible routes for experimental detection.

\section{Preliminaries \& Methodology}
\label{sec:method}

In this section, we present a general overview of the physical framework and outline the methodology used to investigate the acoustic and optical phonon magnetochiral effect (PMCE) in chiral Weyl semimetals. The theoretical formalism proceeds in two main stages.

First, because phonons do not directly couple to magnetic fields, we establish a mechanism through which their interaction with the magnetic field is mediated by chiral electrons. This is accomplished by formulating the Boltzmann kinetic equation (BKE), which is then solved to obtain the nonequilibrium electron distribution in the presence of both phonons and an external magnetic field. In the second stage, this deviation distribution is used to compute the nonequilibrium drag force acting on the lattice. The resulting force enters the elasticity equations, from which we construct the phonon dynamical matrix and determine the corresponding dispersion relations. Before we delve into the technical details of this formalism, we first outline the general physical features of the model.

\begin{figure}[H]
\centering
\includegraphics[width=0.95\columnwidth]{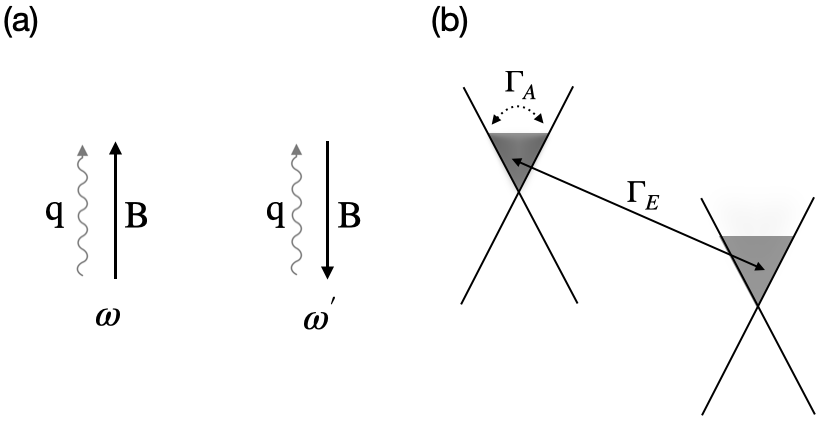}
\caption{(a) Geometry of the phonon magnetochiral effect (PMCE): the relative orientation of the magnetic field $\mathbf{B}$ and the phonon wave vector $\mathbf{q}$ determines the nonreciprocal correction to the phonon dispersion. Reversing the direction of $\mathbf{B}$ (or equivalently $\mathbf{q}$) leads to unequal phonon frequencies, $\omega \neq \omega'$, reflecting the odd-in-$B$ and odd-in-$q$ nature of the effect. (b) Schematic depiction of a chiral Weyl semimetal with two inequivalent Weyl nodes characterized by different Fermi velocities, Fermi energies, and strong Berry curvature and orbital magnetic moments. The quantities $\Gamma_{A}$ and $\Gamma_{E}$ denote the intra-node and inter-node relaxation rates, respectively, which govern local equilibration and chiral charge transfer between Weyl nodes.}
\label{fig:model}
\end{figure}

Figure~\ref{fig:model} summarizes the essential ingredients underlying the phonon magnetochiral effect in chiral Weyl semimetals. As illustrated in Fig.~\ref{fig:model}(a), the PMCE arises only when the phonon wave vector $\mathbf{q}$ has a component parallel to the magnetic field $\mathbf{B}$, leading to a directional (nonreciprocal) shift of the phonon frequency. Because phonons do not couple directly to magnetic fields, this nonreciprocity must originate from their interaction with the electronic system.

In chiral Weyl semimetals, schematically depicted in Fig.~\ref{fig:model}(b), the two Weyl nodes are generically inequivalent—differing in Fermi energy, Fermi velocity, Berry curvature, and orbital magnetic \cite{huang2016, chang2017, tang2017, chang2018, gooth2019, rao2019, sanchez2019, schroter2019, takane2019, senguptapmce}. When a longitudinal phonon propagates along $\mathbf{B}$, these node-specific differences cause the two chiral sectors to respond asymmetrically, generating a nonequilibrium chiral population imbalance governed by the intra-node and inter-node relaxation rates, $\Gamma_A$ and $\Gamma_E$. This imbalance acts back on the lattice as a nonreciprocal drag force, thereby modifying the phonon dispersion in a manner that is odd under $\mathbf{q}\!\to\!-\,\mathbf{q}$ or $\mathbf{B}\!\to\!-\,\mathbf{B}$. This mechanism arising from the interplay of topology, chiral asymmetry, and magnetic field constitutes the microscopic origin of the PMCE which is explored in this work.

Having established the physical setting and the qualitative mechanism behind the PMCE, we now proceed with a detailed formulation of the Boltzmann kinetic equation, which constitutes the first step of our theoretical framework.

\subsection{Boltzmann kinetic equation: Semiclassical formalism}
\label{sec:bke_formalism}

In the presence of an external driving force, the electronic system develops a nonequilibrium response that mediates the coupling between chiral electrons and lattice vibrations. This response is described by the Boltzmann kinetic equation (BKE) for the electronic distribution function $f_{\mathbf{p}}(\mathbf{r},t)$ with collision integral $I_{\rm coll}\{f\}$,
\begin{equation}
\label{eq:bke}
\partial_t f + \dot{\mathbf{r}}\cdot\nabla_{\mathbf{r}} f + \dot{\mathbf{p}}\cdot\nabla_{\mathbf{p}} f = I_{\rm coll}\{f\},
\end{equation}
where the semiclassical equations of motion are given by \cite{Xiao2010, sundaram}
\begin{align}
\label{eq:r}
\dot{\mathbf{r}} &= \nabla_{\mathbf{p}}\varepsilon_{\mathbf{p}}(\mathbf{r},t)
                   + \frac{1}{\hbar}\,\dot{\mathbf{p}} \times \boldsymbol{\Omega}_{\mathbf{p}}, \\
\label{eq:p}
\dot{\mathbf{p}} &= e\,\mathbf{E}(\mathbf{r},t)
                   + e\,\dot{\mathbf{r}} \times \mathbf{B}
                   - \nabla_{\mathbf{r}}\varepsilon_{\mathbf{p}}(\mathbf{r},t),
\end{align}
with $\boldsymbol{\Omega}_{\mathbf{p}}$ the Berry curvature and $\mathbf{E}$ the electric field generated by lattice motion. The electron energy in the presence of a magnetic field and lattice vibrations is
\begin{equation}
\label{eq:electron_energy}
\varepsilon_{\mathbf{p}}(\mathbf{r},t)
 = \varepsilon^{(0)}_{\mathbf{p}} 
 - \mathbf{m}_{\mathbf{p}}\cdot\mathbf{B}
 + \delta\varepsilon(\mathbf{r},t),
\end{equation}
where $\varepsilon^{(0)}_{\mathbf{p}}$ is the band energy in the absence of fields and $\mathbf{m}_{\mathbf{p}}$ is the orbital magnetic moment.

We consider a minimal model of a chiral Weyl semimetal with two Weyl nodes of opposite chirality $\alpha=\pm1$, separated in energy by $2\varpi$ ($\varpi\neq 0$ in a chiral WSM). Following Refs.~\cite{zhong, senguptapmce}, the electronic dispersion, velocity, Berry curvature, and orbital magnetic moment near node $\alpha$ are
\begin{align}
\label{eo}
\varepsilon_0^{(\alpha)}(\mathbf{p}) &= v_F^{(\alpha)} p + \alpha\varpi, \\
\label{v0}
\mathbf{v}^{(\alpha)} &= v_F^{(\alpha)} \hat{\mathbf{p}}, \\
\label{Om}
\boldsymbol{\Omega}^{(\alpha)}_{\mathbf{p}} &= |C| \alpha \hbar^2 \frac{\hat{\mathbf{p}}}{2p^{2}}, \\
\label{m}
\mathbf{m}^{(\alpha)}_{\mathbf{p}} &= -|C|\alpha \hbar \frac{e v_F^{(\alpha)}}{2p}\,\hat{\mathbf{p}}, \\
\label{gradm}
\nabla_{\mathbf{p}}(\mathbf{m}^{(\alpha)}_{\mathbf{p}}\cdot\mathbf{B}) &= 
-|C|\alpha \hbar \frac{e v_F^{(\alpha)}}{2p^2}
\left( \mathbf{B} - 2(\mathbf{B}\cdot\hat{\mathbf{p}})\hat{\mathbf{p}} \right).
\end{align}
Here $v_F^{(\alpha)}$ is the node-dependent Fermi velocity, $|C|=1$ for a single Weyl node (retained as a bookkeeping parameter for geometric contributions), and $\mathbf{p}$ is measured from the Weyl node.

The electronic velocity in a magnetic field, in the absence of lattice motion, is
\begin{equation}
\label{eq:vel}
\tilde{\mathbf{v}}
 = \nabla_{\mathbf{p}}\varepsilon^{(0)}_{\mathbf{p}}
 = \mathbf{v} - \nabla_{\mathbf{p}}(\mathbf{m}_{\mathbf{p}}\cdot\mathbf{B}).
\end{equation}
In zero magnetic field, $\tilde{\mathbf{v}}=\mathbf{v}$. For brevity, the momentum dependence is left implicit and repeated indices are summed over.

The term $\delta\varepsilon$ in Eq.~\eqref{eq:electron_energy} represents the coupling between electrons and phonons. For acoustic modes, the interaction is modeled by node-dependent pseudoscalar deformation potential tensor $\lambda_{ij}^{(\alpha)}\simeq \lambda_1^{(\alpha)}\delta_{ij} + \lambda_2^{(\alpha)} p_i p_j / p^2$ \cite{Kontorovich1984, senguptapmce, falkovsky95}. We retain only the momentum-independent term, such that $\lambda_{ij}^{(\alpha)}\simeq\lambda_1^{(\alpha)}\delta_{ij}$, giving \cite{falkovsky95}
\begin{equation}
\label{eq:acousticenergy}
\delta\varepsilon^{(\text{ac})}(\mathbf{r},t)
 \simeq \lambda_1^{(\alpha)} \delta_{ij}\,u_{ij}(\mathbf{r},t),
\end{equation}
where $u_{ij} = i(q_i u_j + q_j u_i)/2$ is the Fourier transform of the strain tensor and $\mathbf{u}(\mathbf{r},t)$ is the ionic displacement field. The
node-dependent structure of $\lambda_1^{(\alpha)}$ provides the acoustic
analogue of the pseudoscalar electron--phonon coupling, such that $\lambda_{1}^{(+)} \neq \lambda_{1}^{(-)}$.

For the optical phonons considered in this work, we restrict attention to \emph{non-polar} modes. By symmetry, such modes do not induce a change in the macroscopic polarization and therefore possess a vanishing Born effective charge \cite{GonzeLee1997, BornHuang}. As a result, non-polar optical phonons do not generate long-range dipolar fields or LO--TO splitting, and their coupling to magnetic fields is mediated entirely through the electronic subsystem. In this context, the optical-phonon magnetochiral effect arises solely from the Weyl-electron response, in close analogy with the acoustic case.

We model the electron--phonon interaction through the normal-mode expansion of the lattice displacement.
Upon transforming to momentum space, the corresponding energy shift for a
single optical mode takes the form \cite{falkovsky95}
\begin{equation}
\label{eq:opticalenergy}
    \delta\varepsilon^{(\mathrm{op})}_{\alpha}(\mathbf{q},t)
    \simeq g^{(\alpha)}(\mathbf{q})\, \xi_{\mathbf{q}},
\end{equation}
where $\xi_{\mathbf{q}}$ is the optical phonon normal coordinate and
$g^{(\alpha)}$ denotes the electron--phonon coupling at Weyl node $\alpha$. In a chiral Weyl semimetal, symmetry permits the electron--phonon coupling to acquire a pseudoscalar component that changes sign between opposite Weyl
nodes, such that $g^{(+)} \neq g^{(-)}$. Throughout this work we neglect any residual momentum
dependence and treat $g^{(\alpha)}$ as a node-dependent constant.

Using electron-phonon interaction to be linear in $u$ and $\xi$ for the above cases, Eqs.~[\ref{eq:r},\ref{eq:p}] lead to \cite{senguptapmce},
    \begin{widetext}
\begin{align}
  \label{eq:fp}
    \dot{\textbf{p}} &= \frac{e\textbf{E} + e \Tilde{\textbf{v}}\times\textbf{B} + \frac{e^2}{\hbar}\boldsymbol{\Omega}_{\textbf{p}}(\textbf{B}\cdot\textbf{E}) - \partial_{\textbf{r}}\delta\varepsilon+ e \partial_{\textbf{p}}\delta\varepsilon\times\textbf{B} - \frac{e}{\hbar}\boldsymbol{\Omega}_{\textbf{p}}(\textbf{B}\cdot\partial_{\textbf{r}}\delta\varepsilon)}{1+\frac{e}{\hbar}\textbf{B}\cdot\boldsymbol{\Omega}_{\textbf{p}}}\\
  \label{eq:fr}
    \dot{\textbf{r}} &= \frac{\Tilde{\textbf{v}}+ \frac{e}{\hbar} {\bf E}\times {\bf\Omega}_{\bf p} + \frac{e}{\hbar}{\bf B} ({\bf \Omega}_{\bf p}\cdot  \Tilde{\textbf{v}}) + {\partial_{\textbf{p}}\delta\varepsilon} -\frac{1}{\hbar}\partial_{\textbf{r}}\delta\varepsilon\times {\bf \Omega}_{\bf p}+ \frac{e}{\hbar}{\bf B} ({\bf \Omega}_{\bf p}\cdot  {\bf \partial_{\textbf{p}}\delta\varepsilon})  }{1+\frac{e}{\hbar}\textbf{B}\cdot\boldsymbol{\Omega}_{\textbf{p}}}.
\end{align}
\end{widetext}
Next, in Eq.~[\ref{eq:bke}], we seek a solution of the form \cite{senguptapmce},
    \begin{equation}\label{eq:f}
    f_{\bf p}({\bf r},t) = f_{\bf p}^{\rm l. e.}(\textbf{r},t) + \chi_{\bf p}(\textbf{r},t)\frac{\partial f_{0}(\varepsilon_{\textbf{p}}^{(0)})}{\partial \varepsilon_{\textbf{p}}^{(0)}},
    \end{equation}
    with $\chi_{\bf p}(\textbf{r},t)\partial f_{0}(\varepsilon_{\textbf{p}}^{(0)})/\partial \varepsilon_{\textbf{p}}^{(0)}$ being the deviation in electronic distribution function due to presence of phonons and $f_{\bf p}^{\rm l. e.}(\textbf{r},t)$ is the local equilibrium distribution function,  
    \begin{equation}\label{eq:locf}
    \begin{split}
    f_{\bf p}^{\rm l. e.}(\textbf{r},t) \equiv f_{0}(\varepsilon_{\textbf{p}}({\bf r},t)  -\mu_{0} -\delta\mu({\bf r},t))
    \end{split}
    \end{equation}
 with $f_{0}$ being the Fermi-Dirac distribution. The local equilibrium has an effective chemical potential $\mu_0+\delta\mu$, where $\mu_0$ and $\delta\mu$ are the chemical potential terms in the absence and presence of phonons, respectively. Using Eqs.~[\ref{eq:electron_energy} \& \ref{eq:locf}] in Eq.~[\ref{eq:f}] we find
    \begin{equation}\label{eq:ftaylor}
    f \approx f_{0}(\varepsilon^{(0)}_{\bf p}) + (\delta\varepsilon-\delta\mu+\chi)\frac{\partial  f_{0}}{\partial  \varepsilon_{\bf p}^{(0)}}.
    \end{equation}
For the collision integral $I_{\rm coll}$ in Eq.~[\ref{eq:bke}], we use the relaxation time approximation with intra-valley $\Gamma_{A}$ and inter-valley $\Gamma_{E}$ relaxation times, respectively, as shown in Fig.~\ref{fig:model}(b) \cite{zhida2019,senguptapmce},
    \begin{equation}\label{eq:collision}
        \begin{split}
        I_{\rm coll}^{(\alpha)}\{f^{(\alpha)}\} &=-\bigg[\Gamma_{A}\bigg(\chi^{(\alpha)} - \frac{\langle\chi^{(\alpha)}\rangle_{0}}{\nu^{(\alpha)}}\bigg) +\Gamma_{E}\frac{\langle\chi^{(\alpha)}\rangle}{\nu^{(\alpha)}}\bigg]\\
        &\quad \times \frac{\partial f_{0}^{(\alpha)}(\varepsilon_{\bf p}^{(0)})}{\partial \varepsilon_{\bf p}^{(0)}},
        \end{split}
    \end{equation} 
where $\nu^{(\alpha)}$ is the density of states and ${\bf p}$ is the momentum  restricted to the vicinity of Weyl node $\alpha$. The brackets $\langle ... \rangle$ denote an average over the equilibrium (in absence of phonons) Fermi surface, with subscript 0 indicating that the integral is performed at zero magnetic field ($B=0$).

A crucial part of the formalism is to combine Maxwell's equation and the solution of BKE to find  an expression for the electric field induced by lattice vibrations. To do so, we begin by recognizing that for acoustic phonons, the total charge density can be written as $Q+ e n$, where \cite{senguptapmce, Kontorovich1984}

\begin{equation}\label{ion}
    Q = -en_{0}(1-\partial_{\bf r}\cdot{\bf{u}})
\end{equation}
is the ionic charge (to first order in ${\bf u}$, ${\bf{\xi}}$), $n_{0} = \langle\langle f_{0}\rangle\rangle_0$ is the electron density in the absence of lattice vibrations
and
\begin{equation}\label{n}
n =\langle\langle f \rangle \rangle_0
\end{equation}
is the total electron density. Here, the double brackets $\langle\langle ... \rangle\rangle$ denote a momentum integral over the equilibrium ($u=0$, $\xi=0$) Brillouin zone, and the subscript $0$ in $\langle\langle ... \rangle\rangle_0$ is to remind that the integral is done for $B=0$.


In the absence of any external electric field, according to Gauss's law \cite{Kontorovich1984}, 
\begin{equation}\label{gauss}
\epsilon_{lat} \partial_{\bf r}\cdot {\bf E_{0}} = Q+en,
\end{equation}
where $\epsilon_{lat}$ is the high-frequency dielectric permittivity.

Similarly, for the optical phonons, using Gauss's law we have,
\begin{equation}\label{gaussoptical}
\epsilon_{lat} \partial_{\bf r}\cdot {\bf E_{0}} =  -e\langle\langle f_{0}\rangle\rangle_0 + e\langle\langle f\rangle \rangle_0
\end{equation}

We use the above formalism to find the respective electric fields and obtain a solution of the BKE to give us the deviation distribution $\chi$, by using Eqs.~[\ref{eq:fp},\ref{eq:fr}] in Eq.~[\ref{eq:bke}] along with the collision integral from Eq.~[\ref{eq:collision}]. The corresponding expressions in the absence and presence of magnetic field $B$ are given in Secs.~[\ref{subsec:bke_b0} \& \ref{subsec:bke_b}]. Next, we use the expression of the solution of the BKE in the elasticity equations for the acoustic and optical phonons to obtain the drag force and subsequently the phonon dynamical matrices.

\subsection{Elasticity Equations and phonon dispersion relations}

For the acoustic phonons, the elasticity equation for the lattice in the presence of conduction electrons is given by\cite{Kontorovich1984, senguptapmce},

\begin{equation}\label{eq:elasticityacoustic}
\rho \ddot{u}_{h} = \partial_{r_k}\sigma_{hk}^{\rm lat} +\textbf{F}_{h}^{(ac)}({\bf r},t),
\end{equation}

where $h\in\{x,y,z\}$,  $\rho$ is the mass density of the material, $\sigma^{\rm lat}$ is the stress tensor in the absence of conduction electrons,  $\textbf{F}^{(ac)}$ is the drag force exerted by the electrons on the lattice.  The expression of the drag force includes the acoustic electron-phonon interaction and the deviation distribution function from the BKE and is given as\cite{Kontorovich1984,senguptapmce},
\begin{equation}\label{eq:dragforceacoustic}
\begin{split}
F_{h}^{(ac)} &= \partial_{r_k} \langle\!\langle \lambda_{hk} f\rangle\!\rangle.\\
\end{split}
\end{equation}
In Eq.~[\ref{eq:elasticityacoustic}], we have neglected terms involving the electronic current density ($\textbf{j}_{\rm el}$) and ionic current density ($\textbf{j}_{\rm lat}$) as the contribution from these terms are negligible compared to that of the drag force\cite{senguptapmce}. The stress tensor is related to strain through $\sigma_{hk}^{\rm lat} = s_{hkim} u_{im}$, where $s_{hkim}$ is the stiffness tensor whose general form depends on the crystal symmetry of the material. 

Similarly, for the optical phonons, the elasticity equation for the lattice in the presence of conduction electrons is given by\cite{falkovsky95, rinkel2017, rinkel2019, song2016, Kontorovich1984}

\begin{equation}\label{eq:elasticityoptic}
M (\omega^{2} - \omega_{0}^{2}) \xi = \textbf{F}^{(op)}({\bf r},t),
\end{equation}

Here, the drag force $F^{(op)}$ relating the deviation distribution from BKE and electron-optical phonon coupling for a sample volume $\mathcal{V}$ and $N$ number of unit cells in the crystal is given as, \cite{Kontorovich1984,rinkel2019,falkovsky95}
\begin{equation}\label{eq:dragforceoptic}
F^{(op)} = \frac{\mathcal{V}}{N}\langle\langle g f \rangle\rangle.
\end{equation}

In Secs.~[\ref{subsec:drag_force_b0_acoustic},  \ref{subsec:drag_force_b_acoustic}, \ref{subsec:drag_force_b0_optical} \& \ref{subsec:drag_force_b_optical}], we show the detailed derivations for the drag force and dispersion relations in absence and presence of magnetic field $B$, respectively for both acoustic and optical phonons.

In what follows, we apply the foregoing formalism to the zero magnetic field
case and derive the chiral electron–phonon renormalization of both acoustic and
optical phonon modes.
    
\section{Chiral electron effects on phonon dispersion at zero magnetic field}
\label{sec:zero-mag}

In this section, we apply the above formalism to solve the BKE in the absence of a
magnetic field and derive the resulting drag force exerted by electrons on the
lattice, along with the corresponding phonon dispersion relations. The analysis is
carried out for both acoustic and optical modes. We begin by solving the Boltzmann
equation for the electron response to the respective lattice vibrations.

\subsection{Solution of the Boltzmann equation: Deviation distribution for $\mathbf{B}=0$}
\label{subsec:bke_b0}

\subsubsection{Electron-Acoustic Phonon Interactions}
\label{subsec:acoustic_bke0}
 We plug the electron-acoustic phonon interaction given by Eq.~[\ref{eq:acousticenergy}] in Eqs.~[\ref{eq:fp}, \ref{eq:fr}], and subsequently use Eqs. ~[\ref{eq:bke}, \ref{eq:collision}] to derive the BKE,

\begin{widetext}
\begin{align}\label{BKEacoustic}
    &-i\omega \bigg(1+ \frac{e}{\hbar} {\bf{B\cdot\Omega^{(\alpha)}_{\textbf{p}}}}\bigg)\chi^{(\alpha)} + i\left( {\textbf{q}\cdot\Tilde{\textbf{v}}^{(\alpha)}} + \frac{e}{\hbar}(\textbf{q}\cdot\textbf{B})(\boldsymbol{\Omega}^{(\alpha)}\cdot\Tilde{\textbf{v}}^{(\alpha)})\right)\chi^{(\alpha)} + e(\Tilde{\textbf{v}}^{(\alpha)}\times\textbf{B})\cdot\partial_{\bf p}\chi^{(\alpha)}+ \bigg(1+ \frac{e}{\hbar} {\bf{B\cdot\Omega^{(\alpha)}_{\textbf{p}}}}
    \bigg)\bigg[\Gamma_{A}\chi^{(\alpha)}\nonumber\\
     & -\Gamma_{A}\frac{\langle\chi^{(\alpha)}\rangle}{\nu^{(\alpha)}} + \Gamma_{E}\frac{\langle\chi^{(\alpha)}\rangle}{\nu^{(\alpha)}}\bigg] = -\bigg(1+ \frac{e}{\hbar} {\bf{B\cdot\Omega^{(\alpha)}_{\textbf{p}}}}\bigg)\bigg[-i\omega\lambda_{i j}^{(\alpha)} u_{i j} +i\omega\delta\mu\bigg] 
    + i \delta\mu \left( {\textbf{q}\cdot\Tilde{\textbf{v}}^{(\alpha)}} + \frac{e}{\hbar}(\textbf{q}\cdot\textbf{B})(\boldsymbol{\Omega}^{(\alpha)}\cdot\Tilde{\textbf{v}}^{(\alpha)})\right)\nonumber\\ &- \Tilde{\textbf{v}}^{(\alpha)}\cdot\left(e\textbf{E} + \frac{e^2}{\hbar}\boldsymbol{\Omega}^{(\alpha)}(\textbf{B}\cdot\textbf{E}) \right)
    - i\omega e(\Tilde{\textbf{v}}^{(\alpha)}\times\textbf{B})\cdot{\textbf{u}},
\end{align}
\end{widetext}
where, we have used $\partial_t\rightarrow -i\omega$, $\partial_{\bf r}\rightarrow i\textbf{q}$ and $\chi_{\bf p}(\textbf{r},t)\rightarrow \chi_{\bf p}(\textbf{q},\omega)$, and the complex number $i$ should not be confused with the subscript $i$ appearing in $\lambda_{i j}$, $u_{i j}$, $u_i$.
Also, $u_{i j} = i (q_i u_j + q_j u_i)/2$ is the Fourier transform of the strain tensor \cite{senguptapmce}. 

For the zeroth order in magnetic field, the above equation reduces to
\begin{equation}\label{BKEacousticB0}
\begin{split}
    &(-i\omega + i\textbf{q}\cdot\textbf{v}^{(\alpha)}+\Gamma_{A})\chi_{0}^{(\alpha)} - (\Gamma_{A}-\Gamma_E)\frac{\langle\chi_{0}^{(\alpha)}\rangle_0}{\nu^{(\alpha)}_0} \\
    &=  -\lambda_{ij}^{(\alpha)}\omega q_{j}u_{i}- i\omega\delta\mu_{0} - e{\bf E}_0\cdot{\bf v}^{(\alpha)}+ i\textbf{q}\cdot{\textbf{v}^{(\alpha)}}\delta\mu_{0},\\
\end{split}
\end{equation}
where the subscript $0$ in $\langle ... \rangle_0$ indicates that the average is taken on the $B=0$ Fermi surface and
\begin{equation}\label{DOSFS}
\nu_{0}^{(\alpha)} = \frac{1}{(2\pi\hbar)^{3}}\int\frac{dS^{(\alpha)}_{F}}{v^{(\alpha)}_{F}},
\end{equation}
with $(\theta,\phi)$ as the polar and azimuthal angles in spherical coordinates,
\begin{equation}\label{FS}
    dS^{(\alpha)}_{F} = (p^{(\alpha)}_{F})^{2} \sin\theta d\theta d\phi
\end{equation}
is the surface area element on the Fermi surface near node $\alpha$. 

Using Gauss's law from Eq.~[\ref{gauss}], we arrive at

\begin{equation}\label{acousticdelmu0}
\delta{\mu}_{0} = \frac{i\textbf{q}\cdot \textbf{E}_{0} \epsilon_{lat}}{ e\nu_0}  + \frac{\langle\lambda_{ij}\rangle_0}{\nu_{0}}iq_{j}u_{i}.
\end{equation}
Plugging the above equation in Eq.~[\ref{BKEacousticB0}], we find
\begin{widetext}
\begin{equation}\label{acousticchi0B0}
\begin{split}
  \chi_{0}^{(\alpha)} &= R^{(\alpha)} \bigg[ \bigg(-\lambda_{ij}^{(\alpha)} + \frac{\langle\lambda_{ij}\rangle_0}{\nu_{0}}\bigg)\omega q_{j}u_{i} - ({\textbf{q}}\cdot{\textbf{v}^{(\alpha)}})\frac{\langle\lambda_{ij}\rangle_0}{\nu_{0}}q_{j}u_{i}
  - e\textbf{E}_{0}\cdot{\textbf{v}^{(\alpha)}} + \frac{\omega (\textbf{q}\cdot\textbf{E}_{0})\epsilon_{lat}}{e\nu_0}
  -\frac{(\textbf{q}\cdot{\textbf{v}^{(\alpha)})(\textbf{q}\cdot\textbf{E}_{0})}\epsilon_{lat}}{e\nu_0} \\
&+ (\Gamma_{A}-\Gamma_E)\frac{\langle\chi_{0}^{(\alpha)}\rangle_0}{\nu^{(\alpha)}_0} \bigg],
\end{split}
\end{equation}
\end{widetext}
where,
\begin{equation}\label{Ralpha}
  R^{(\alpha)} = (-i\omega + i{\textbf{q}}\cdot{\textbf{v}^{(\alpha)}} +\Gamma_{A})^{-1}.
\end{equation}
Before turning to the derivation of the electric field, we summarize several useful
identities that will be employed in the following subsections. For long-wavelength
acoustic phonons, where the phonon momentum typically satisfies
$q \lesssim 5\times 10^{5}\,{\rm m^{-1}}$, the associated energy scales are
$\hbar v_{F} q \lesssim 10^{-2}\,{\rm meV}$ and
$\hbar\omega \lesssim 10^{-4}\,{\rm meV}$. Both are much smaller than
$\hbar\Gamma_{A}$ in moderately disordered Weyl semimetals, where one expects
$\hbar\Gamma_{A} \sim 1$--$10\,{\rm meV}$ at low temperatures. In contrast,
the internode relaxation rate is significantly slower:
$\hbar\Gamma_{E} \sim 10^{-2}\,{\rm meV}$, reflecting the suppression of
intervalley impurity scattering relative to intravalley processes. The hierarchy
$\Gamma_{A} \gg \Gamma_{E}$, together with
$\Gamma_{A} \gg v_{F}^{(\alpha)} q \gg \omega$, will be used extensively in the
simplifications that follow. Under these conditions, Taylor expanding
Eq.~\eqref{Ralpha} yields

\begin{equation}\label{Racoustic0}
\begin{split}
  R^{(\alpha)} &=\frac{1}{\Gamma_{A}}\bigg[1- i\bigg\{\frac{{\textbf{q}}\cdot{\textbf{v}^{(\alpha)}}}{\Gamma_{A}}\bigg\} +2 \bigg\{\bigg(\frac{{\textbf{q}}\cdot{\textbf{v}^{(\alpha)}}}{\Gamma_{A}}\bigg)\bigg(\frac{\omega}{\Gamma_{A}}\bigg)\bigg\}\\
  &\quad- \bigg\{\bigg(\frac{{\textbf{q}}\cdot{\textbf{v}^{(\alpha)}}}{\Gamma_{A}}\bigg)^{2}\bigg\}+ \bigg\{\frac{i\omega}{\Gamma_{A}}\bigg\} -\bigg\{\bigg(\frac{\omega}{\Gamma_{A}}\bigg)^{2}\bigg\}\bigg]\\
\end{split}
\end{equation}
Taking Fermi Surface (FS) average we get,
\begin{equation}\label{Rusec}
  \langle R^{(\alpha)}\rangle_0\simeq \frac{\nu_{0}^{(\alpha)}}{\Gamma_A}\left(1+\frac{i\omega}{\Gamma_A} \right),
\end{equation}
where we have used $(qv_{F}^{(\alpha)}/\Gamma_{A})^{2}\ll 1 $ and $(\omega/\Gamma_{A})^{2}\ll 1$.
Using Eq.~[\ref{Racoustic0}], we find other important identities,
\begin{equation}\label{identities}
\begin{split}
\langle (\hat{\textbf{q}}\cdot \textbf{v}^{(\alpha)}) R^{(\alpha)}\rangle_{0}  &\approx - iq D^{(\alpha)}\frac{\nu_{0}^{(\alpha)}}{\Gamma_{A}} \bigg( 1
+ \frac{2i\omega }{\Gamma_{A}}\bigg) \\
\langle (\hat{\textbf{q}}\cdot \textbf{v}^{(\alpha)})^{2} R^{(\alpha)}\rangle_{0}
&\approx \nu_{0}^{(\alpha)} D^{(\alpha)}\bigg(1 + \frac{i\omega}{\Gamma_{A}} \bigg) \\
\end{split}
\end{equation}
where, we have used the diffusion constant $D$,
\begin{equation}\label{Diff}
D^{(\alpha)} = \frac{1}{3}\frac{(v_{F}^{(\alpha)})^{2}}{\Gamma_{A}}
\end{equation}
In the above equations, we have neglected terms that appear in the order of
\begin{equation}\label{termq}
\frac{(qv_{F}^{(\alpha)})^{2}}{\Gamma_{A}^{2}} \sim 10^{-6} \ll 1.
\end{equation}

Next we calculate the electric field, by taking the Fermi surface average of the Eq.~[\ref{acousticchi0B0}] and impose normalization condition $\sum_{\alpha}\langle \chi^{(\alpha)}\rangle = 0$ \cite{Kontorovich_1984, senguptapmce} and use the identities given by Eq.~[\ref{identities}], such that the longitudinal component of the electric field due to electron-acoustic phonon interactions reduces to \cite{antebi}
\begin{equation}\label{E0rudra}
eE_{0,\parallel} =-q\frac{\langle \lambda_{ij}\rangle_{0}}{\nu_{0}} q_{j} u_{i}.
\end{equation}
The resulting electrochemical potential difference between the two nodes, which encodes the chiral charge imbalance, takes the form:
\begin{equation}\label{avchi0}
\begin{split}
\langle\chi_{0}^{(+)}\rangle_{0} &= -\langle\chi_{0}^{(-)}\rangle_{0} =-\frac{\omega q_{j}u_{i}}{\Gamma_{E}\nu_{0}}\bigg(1 + \frac{i\omega}{\Gamma_{E}}\bigg)\nu_{0}^{(+)}\nu_{0}^{(-)}\\
&\times(\lambda_{1}^{(+)} - \lambda_{1}^{(-)})\delta_{ij}.\\
\end{split}
\end{equation}


Invoking the limit ($\kappa^{2} = e^{2}\langle1\rangle/\epsilon_{lat})\rightarrow \infty$ in Eq.~[\ref{acousticdelmu0}] we find,

\begin{equation}\label{FTdelmunewe0}
\delta{\mu}_{0} \approx  \frac{\langle\lambda_{ij}\rangle_0}{\nu_0}iq_{j}u_{i}.\\
\end{equation}

We plug the expressions from Eq.~(\ref{E0rudra}) and Eq.~(\ref{FTdelmunewe0}) in Eq.~(\ref{acousticchi0B0}), to get

\begin{equation}\label{chi0B0E0new1}
\begin{split}
  \chi_{0}^{(\alpha)} &= R^{(\alpha)} \bigg[ \bigg(\frac{\langle\lambda_{ij}\rangle_0}{\nu_0}-\lambda_{ij}^{(\alpha)} \bigg)\omega q_{j}u_{i}
  + (\Gamma_{A}-\Gamma_E)\frac{\langle\chi_{0}^{(\alpha)}\rangle_0}{\nu^{(\alpha)}_0} \bigg],\\
\end{split}
\end{equation}

In Subsection~\ref{subsec:acoustic_bke_b}, where we analyze the magnetic-field
induced corrections, we will use the above expressions for
$\chi_{0}^{(\alpha)}$, $E_{0}$, $\delta\mu_{0}$, and
$\langle\chi_{0}^{(\alpha)}\rangle$ to obtain the deviation distribution relevant
to the electron–acoustic phonon interaction. In the next subsection, we provide similar expressions for the interaction of chiral electrons with optical phonons.

\subsubsection{Electron-Optical Phonon Interaction}
\label{subsec:electron-optical_bke0}

As in the previous subsection, we first solve the BKE in the absence of a magnetic
field and then use this solution within a perturbative treatment to obtain the
deviation distribution in the presence of a magnetic field, as described in
Subsection~\ref{subsec:electron-optical_bke_b}. Using
Eq.~\eqref{eq:opticalenergy} together with Eqs.~\eqref{eq:bke},
\eqref{eq:collision}, \eqref{eq:fp}, and \eqref{eq:fr}, the Fourier-transformed BKE
for the optical-phonon–electron interaction takes the form:
\begin{widetext}
\begin{align}\label{BKEoptical}
    &-i\omega \bigg(1+ \frac{e}{\hbar} {\bf{B\cdot\Omega^{(\alpha)}_{\textbf{p}}}}\bigg)\chi^{(\alpha)} +\bigg(1+ \frac{e}{\hbar} {\bf{B\cdot\Omega^{(\alpha)}_{\textbf{p}}}}
    \bigg)\bigg[\Gamma_{A}\chi^{(\alpha)}-\Gamma_{A}\frac{\langle\chi^{(\alpha)}\rangle}{\nu^{(\alpha)}} + \Gamma_{E}\frac{\langle\chi^{(\alpha)}\rangle}{\nu^{(\alpha)}}\bigg] = \bigg(1+ \frac{e}{\hbar} {\bf{B\cdot\Omega^{(\alpha)}_{\textbf{p}}}}\bigg)\bigg[i\omega g^{(\alpha)}\xi_{{\textbf{q}}}  -i\omega\delta\mu\bigg]\nonumber \\
    &+ i (\delta\mu - \chi^{(\alpha)}) \left( {\textbf{q}\cdot\Tilde{\textbf{v}}^{(\alpha)}} + \frac{e}{\hbar}(\textbf{q}\cdot\textbf{B})(\boldsymbol{\Omega}^{(\alpha)}\cdot\Tilde{\textbf{v}}^{(\alpha)})\right)- \Tilde{\textbf{v}}^{(\alpha)}\cdot\left(e\textbf{E} + \frac{e^2}{\hbar}\boldsymbol{\Omega}^{(\alpha)}(\textbf{B}\cdot\textbf{E}) \right)
\end{align}
\end{widetext}
We note that the difference from the acoustic case is observed only as is expected in the electron-phonon coupling terms. In the absence of magnetic field, the above equation reduces to, 
\begin{equation}\label{BKEopticalB0}
\begin{split}
    &(-i\omega + i\textbf{q}\cdot\textbf{v}^{(\alpha)}+\Gamma_{A})\chi_{0}^{(\alpha)} - (\Gamma_{A}-\Gamma_E)\frac{\langle\chi_{0}^{(\alpha)}\rangle_0}{\nu^{(\alpha)}_0} \\
    &=  i\omega(g^{(\alpha)}\xi_{{\textbf{q}}}- \delta\mu_{0}) - e{\bf E}_0\cdot{\bf v}^{(\alpha)}+ i\textbf{q}\cdot{\textbf{v}^{(\alpha)}}\delta\mu_{0}.\\
\end{split}
\end{equation}
Following Gauss's law from Eq.~[\ref{gauss}], we find
\begin{equation}\label{opticaldelmu0}
\delta{\mu}_{0} = \frac{i\textbf{q}\cdot \textbf{E}_{0} \epsilon_{lat}}{ e\nu_0}  + \frac{\langle g\rangle_0}{\nu_{0}}\xi_{{\textbf{q}}},
\end{equation}
which we then plug in Eq.~[\ref{BKEopticalB0}], to get

\begin{equation}\label{opticchi0B0}
\begin{split}
  \chi_{0}^{(\alpha)} &= R^{(\alpha)} \bigg[i\omega \bigg(g^{(\alpha)} - \frac{\langle g\rangle_0}{\nu_{0}}\bigg) \xi_{{\textbf{q}}} + i ({\textbf{q}}\cdot{\textbf{v}^{(\alpha)}})\frac{\langle g\rangle_0}{\nu_{0}}\xi_{{\textbf{q}}}\\
  &- e\textbf{E}_{0}\cdot{\textbf{v}^{(\alpha)}} + \frac{\omega (\textbf{q}\cdot\textbf{E}_{0})\epsilon_{lat}}{e\nu_0}
  -\frac{(\textbf{q}\cdot{\textbf{v}^{(\alpha)})(\textbf{q}\cdot\textbf{E}_{0})}\epsilon_{lat}}{e\nu_0} \\
&+ (\Gamma_{A}-\Gamma_E)\frac{\langle\chi_{0}^{(\alpha)}\rangle_0}{\nu^{(\alpha)}_0} \bigg],\\
\end{split}
\end{equation}

where, $R^{(\alpha)}$ is given by 
\begin{equation}\label{Roptic0}
\begin{split}
  R^{(\alpha)} &=\frac{i}{\omega}\bigg[1+ \bigg\{\frac{{\textbf{q}}\cdot{\textbf{v}^{(\alpha)}}}{\omega}\bigg\} -2 \bigg\{\bigg(\frac{{\textbf{q}}\cdot{\textbf{v}^{(\alpha)}}}{\omega}\bigg)\bigg(\frac{i\Gamma_{A}}{\omega}\bigg)\bigg\}\\
  &\quad+ \bigg\{\bigg(\frac{{\textbf{q}}\cdot{\textbf{v}^{(\alpha)}}}{\omega}\bigg)^{2}\bigg\}- \bigg\{\frac{i\Gamma_{A}}{\omega}\bigg\} -\bigg\{\bigg(\frac{\Gamma_{A}}{\omega}\bigg)^{2}\bigg\}\bigg]\\
\end{split}
\end{equation}  

Before we proceed to the derivation of the electric field, let us mention some important identities which we will use in the following subsections. In the optical phonon case, our theory adheres to the following conditions $\omega \gg v_F^{(\alpha)} q \gg \Gamma_A $, such that taking the FS average of Eq.~[\ref{Roptic0}], we get

\begin{equation}\label{Roptic}
  \langle R^{(\alpha)}\rangle_0\simeq \frac{i\nu_{0}^{(\alpha)}}{\omega}\left(1-\frac{i\Gamma_A}{\omega} \right).
\end{equation}

Using the above, we find
\begin{equation}\label{identitiesoptic}
\begin{split}
\langle (\hat{\textbf{q}}\cdot \textbf{v}^{(\alpha)}) R^{(\alpha)}\rangle_{0}  &\approx  \frac{i\nu_{0}^{(\alpha)}q D^{(\alpha)}\Gamma_{A}}{\omega^{2}} \bigg( 1
- \frac{2i\Gamma_{A}}{\omega}\bigg), \\
\end{split}
\end{equation}
where, $D^{(\alpha)}$ is given by Eq.~[\ref{Diff}].
In the above equations, we have neglected terms that appear in the order of
\begin{equation}\label{termq}
\frac{\Gamma_{A}^{2}}{\omega^{2}} \ll 1 , \frac{(qv_{F}^{(\alpha)})^{2}}{\omega^{2}} \sim 10^{-6} \ll 1.
\end{equation}
when compared with leading order real terms. 

To calculate the electric field, we take the Fermi surface average of the Eq.~[\ref{opticchi0B0}] and impose normalization condition $\sum_{\alpha}\langle \chi^{(\alpha)}\rangle = 0$\cite{Kontorovich_1984, senguptapmce} and the above identities given by Eq.~[\ref{identitiesoptic}], such that the longitudinal electric field due to electron-optical phonon interaction reduces to
\begin{equation}\label{E0rudraoptical}
eE_{0,\parallel} =iq\frac{\langle g\rangle_{0}}{\nu_{0}}\xi_{{\textbf{q}}} ,
\end{equation}
and the chiral charge imbalance generated takes up the form:
\begin{equation}\label{avchi0optical}
\begin{split}
\langle\chi_{0}^{(+)}\rangle_{0} &= -\langle\chi_{0}^{(-)}\rangle_{0} =-\frac{ \xi_{{\textbf{q}}}}{\nu_{0}}\bigg(1 - \frac{i\Gamma_{E}}{\omega}\bigg)\nu_{0}^{(+)}\nu_{0}^{(-)}\\
&\times(g^{(+)} - g^{(-)}).\\
\end{split}
\end{equation}
Using the expression of $E_{0,\parallel}$ in Eq.~[\ref{opticaldelmu0}] and invoking the limit $\kappa^{2}\rightarrow \infty$, 
we find
\begin{equation}\label{FTdelmunewe0optical}
\delta{\mu}_{0} \approx  \frac{\langle g\rangle_0}{\nu_0} \xi_{{\textbf{q}}}.\\
\end{equation}

We plug the expressions from Eq.~[\ref{E0rudraoptical}] \& Eq.~[\ref{FTdelmunewe0optical}] in Eq.~(\ref{opticchi0B0}) and neglect terms of order $q^{2}/\kappa^{2} \ll 1$, to get

\begin{equation}\label{chi0B0E0new1}
\begin{split}
  \chi_{0}^{(\alpha)} &= R^{(\alpha)} \bigg[i\omega \bigg(g^{(\alpha)} - \frac{\langle g\rangle_0}{\nu_{0}}\bigg)\xi_{{\textbf{q}}} 
  + (\Gamma_{A}-\Gamma_E)\frac{\langle\chi_{0}^{(\alpha)}\rangle_0}{\nu^{(\alpha)}_0} \bigg].\\
\end{split}
\end{equation}
In the Subsection~[\ref{subsec:electron-optical_bke_b}], we will use the expression of $\chi_{0}^{(\alpha)}$, $\delta\mu_{0}$, $E_{0,\parallel}$ and $\langle\chi_{0}^{(\alpha)}\rangle_{0}$ to find the solution of BKE in presence of magnetic field.

Next, we derive the acoustic and optical phonon dispersion relations at zero
magnetic field using the expressions obtained in the preceding subsections.

\subsection{Acoustic dispersion derived from drag force and elasticity equations}
\label{subsec:drag_force_b0_acoustic}

In this subsection, we examine the effect of the deviation distribution on the
longitudinal acoustic mode. We begin with the elasticity equation given by  Eq.~\eqref{eq:elasticityacoustic}, which requires the evaluation of the drag force
$F^{(\mathrm{ac})}$ exerted by the conduction electrons. We first present the
expressions for $F^{(\mathrm{ac})}$ and for the elasticity equation in the absence of
a magnetic field, and then extend the analysis to include the magnetic-field
corrections, treated to linear order in $B$, in Subsection~\ref{subsec:drag_force_b_acoustic}.

The expression of the drag force includes the acoustic electron-phonon interaction and the deviation distribution function given by\cite{Kontorovich_1984,senguptapmce},
\begin{equation}\label{dragforceacoustic}
\begin{split}
F_{h}^{(ac)} &= \partial_{r_k} \langle\!\langle \lambda_{hk} f\rangle\!\rangle\\
& \approx -i q_k \sum_{\alpha=+,-}\left(\langle\lambda^{(\alpha)}_{h k} \chi^{(\alpha)}\rangle_0 - \delta\mu\langle\lambda^{(\alpha)}_{h k}\rangle_0\right)
\end{split}
\end{equation}
where, we have used Eq.~[\ref{eq:ftaylor}]. In the absence of magnetic field, we use the expression for $\chi_{0}^{(\alpha)}$ from Eq.~[\ref{avchi0}] and retain the leading order contribution to get, 
\begin{equation}\label{dragacousticB0}
\begin{split}
    F_{z}^{(ac)}(B=0) = i\omega q_{z}^{2}u_{z}(\lambda_{1}^{(+)} - \lambda_{1}^{(-)})^{2}\bigg(\frac{1}{\Gamma_{E}} + \frac{i\omega}{\Gamma_{E}^{2}}\bigg) \frac{\nu_{0}^{(+)}\nu_{0}^{(-)}}{\nu_{0}}.
\end{split}
\end{equation}

Using the above in Eq.~[\ref{eq:elasticityacoustic}], we find an expression of elasticity equation in the absence of magnetic field,
\begin{equation}\label{elasticityacousticB0}
    \rho\omega^{2}u_{z} = q_{z}^{2}s_{33} u_{z} + i\omega q_{z}^{2}u_{z}(\lambda_{1}^{(+)} - \lambda_{1}^{(-)})^{2}\bigg(\frac{1}{\Gamma_{E}} + \frac{i\omega}{\Gamma_{E}^{2}}\bigg) \frac{\nu_{0}^{(+)}\nu_{0}^{(-)}}{\nu_{0}},
\end{equation}

which leads to the following expression for the longitudinal acoustic phonon dispersion relation,

\begin{equation}
\begin{split}
\label{eq:long-ac-b0}
0 &= q_{z}^{2}c_{s}^{2} - \omega^{2}\bigg[1 + \frac{q_{z}^{2}(\lambda_{1}^{(+)} - \lambda_{1}^{(-)})^{2}}{\rho\Gamma_{E}^{2}}  \frac{\nu_{0}^{(+)}\nu_{0}^{(-)}}{\nu_{0}}\bigg]\\
& + \omega\bigg[\frac{iq_{z}^{2}(\lambda_{1}^{(+)} - \lambda_{1}^{(-)})^{2}}{\rho\Gamma_{E}}  \frac{\nu_{0}^{(+)}\nu_{0}^{(-)}}{\nu_{0}}\bigg],\\
\end{split}
\end{equation}
where, the speed of sound is $c_{s}(0) = \sqrt{s_{33}/\rho}$ = $2\times10^{3}$ m/s. We show the numerical results for the acoustic phonon dispersion at $B=0$ in Fig. ~[\ref{fig:real_acoustic}]

\subsection{Optical dispersion derived from drag force and elasticity equations}
\label{subsec:drag_force_b0_optical}

The elasticity equation for the optical phonons follows from Eq.~[\ref{eq:elasticityoptic}]. The starting point is the calculation of the drag force $F^{(op)}$ relating the deviation distribution and electron-optical phonon coupling for a sample volume $\mathcal{V}$ and $N$ number of unit cells in the crystal \cite{Kontorovich_1984,rinkel2019,falkovsky95},
\begin{equation}\label{dragforceoptic}
F^{(op)} = \frac{\mathcal{V}}{N}\langle\langle g f \rangle\rangle \simeq \frac{\mathcal{V}}{N}\sum_{\alpha =+,-} g^{(\alpha)}
\langle \chi^{(\alpha)}\rangle_{0}.
\end{equation}
In the absence of magnetic field, we use Eq.~[\ref{avchi0optical}], to find

\begin{equation}\label{FopticalB0}
F^{(op)} (B=0) = -\frac{\mathcal{V}}{N}(g^{(+)} - g^{(-)})^{2}\bigg(1 - \frac{i\Gamma_{E}}{\omega}\bigg)\frac{\nu_{0}^{(+)}\nu_{0}^{(-)}}{\nu_{0}}\xi_{{\textbf{q}}}.
\end{equation}
Plugging the above in Eq.~[\ref{eq:elasticityoptic}], we write the elasticity equation in absence of magnetic field,
\begin{equation}\label{elasticityopticB0}
(\omega^{2} -\omega_{0}^{2}) \xi_{{\textbf{q}}} = -\frac{\mathcal{V}}{NM} (g^{(+)} - g^{(-)})^{2}\bigg(1 -\frac{i\Gamma_{E}}{\omega}\bigg)\frac{\nu_{0}^{(+)}\nu_{0}^{(-)}}{\nu_{0}} \xi_{{\textbf{q}}}\\
\end{equation}
In the long-wavelength approximation, let us introduce the factor \cite{rinkel2019}:
\begin{equation}\label{eq:optical_factor}
(g^{(+)} -g^{(-)})^{2} \mathcal{V}/NM = \Delta^{2}/\rho' a^{2},
\end{equation}
where $\Delta$ is the deformation potential produced by the pseudoscalar optical phonon in energy units, $\rho' = NM/\mathcal{V}$ is the mass density of the crystal and $a$ is the typical linear dimension of the unit cell \cite{rinkel2019}. Eq.~[\ref{elasticityopticB0}] thus reduces to,
\begin{equation}\label{elasticityopticB0def}
(\omega^{2} -\omega_{0}^{2}) \xi_{{\textbf{q}}} = -\frac{\Delta^{2}}{\rho' a^{2}}\bigg(1 -\frac{i\Gamma_{E}}{\omega}\bigg)\frac{\nu_{0}^{(+)}\nu_{0}^{(-)}}{\nu_{0}} \xi_{{\textbf{q}}}.
\end{equation}

The numerical results for the optical phonon dispersion is shown in Fig.~[\ref{fig:real_optical}] for specific material parameters. In the next section, we now proceed to show the derivations for the magnetic field dependent corrections to the phonon dispersion relations.

\section{Chiral electron effects on phonon dispersion in the presence of a magnetic field}
\label{sec:pmce}

In this section, we extend the formalism developed in Sec.~\ref{sec:method} to the
case where a magnetic field $B$ is applied. The magnetic field modifies both the
deviation distribution and the resulting drag force, giving rise to nonreciprocal
corrections in the phonon dispersion relations. We begin by solving the Boltzmann
equation in the presence of a magnetic field, treating the field to linear order.

\subsection{Deviation distribution in presence of magnetic field: at $\textbf{B} \ne 0$}
\label{subsec:bke_b}

In this subsection, we obtain the solution of the BKE in the presence of a magnetic
field for both acoustic and optical phonons. We begin with the case of
electron–acoustic phonon coupling.

\subsubsection{Electron-Acoustic Phonon Interactions}
\label{subsec:acoustic_bke_b}

Starting from Eq.~[\ref{BKEacoustic}], we apply the following perturbative approximations to linear order in $B$, $\chi= \chi_0 + \chi_1,$ $\delta\mu= \delta\mu_0 + \delta\mu_1$, $\textbf{E}= \textbf{E}_0 + \textbf{E}_1$, where the subscripts $0$ and $1$ have the meanings of zeroth order and linear order in $B$, respectively. Next, we collect terms that are linear order in $B$, such that the expression for BKE reduces to
\begin{widetext}
\begin{align}\label{BKEacousticB1}
   & \frac{\chi_{1}^{(\alpha)}}{R^{(\alpha)}}= -\frac{e}{\hbar}(\textbf{B}\cdot\boldsymbol{\Omega}^{(\alpha)})\bigg[\bigg(\lambda_{ij}^{(\alpha)} -\frac{\langle\lambda_{ij}\rangle_0}{\nu_0}\bigg)\omega q_{j}u_{i} + \chi_{0}^{(\alpha)}(\Gamma_{A}-i\omega) - (\Gamma_{A}-\Gamma_E)\frac{\langle\chi_{0}^{(\alpha)}\rangle_0}{\nu^{(\alpha)}_0}\bigg]\nonumber \\
  &+ i\textbf{q}\cdot\bigg(\frac{e}{\hbar}(\boldsymbol{\Omega}^{(\alpha)}\cdot\textbf{v}^{(\alpha)})\textbf{B} -\partial_{\textbf{p}}(\textbf{m}^{(\alpha)}\cdot\textbf{B}) \bigg)\bigg(iq_{j}u_{i}\frac{\langle\lambda_{ij}\rangle_0}{\langle1\rangle_0} -\chi_{0}^{(\alpha)}\bigg)-e(\textbf{v}^{(\alpha)}\times\textbf{B})\cdot(i\omega \textbf{u}+\partial_{\bf p}\chi_0^{(\alpha)})\nonumber \\
  &+ i({\textbf{q}}\cdot{\textbf{v}^{(\alpha)}}-\omega)\delta\mu_{1}+ e\textbf{E}_{0}\cdot\left( \partial_{\textbf{p}}(\textbf{m}^{(\alpha)}\cdot\textbf{B}) - \frac{e}{\hbar}(\boldsymbol{\Omega}_{\textbf{p}}^{(\alpha)}\cdot\textbf{v}^{(\alpha)})\textbf{B} \right)- e\textbf{v}^{(\alpha)}\cdot\textbf{E}_1 + (\Gamma_{A}-\Gamma_E)\frac{\langle\chi_{1}^{(\alpha)}\rangle_0}{\nu^{(\alpha)}_0},
\end{align}
\end{widetext}
In the above equation, we have three unknowns $\chi_{1}^{(\alpha)}$, $\delta\mu_{1}$ and $E_{1}$, and following a similar procedure as given in the above Secs.~[\ref{sec:method} \& \ref{subsec:bke_b0}], we find these unknowns via Gauss's Law and normalization condition. In the presence of magnetic field, using Gauss's law we find,

\begin{equation}\label{deltamu1acoustic}
\delta\mu_{1} = \frac{i\textbf{q}\cdot\textbf{E}_{1} \epsilon_{lat}}{e\nu_0}.
\end{equation}
We then plug the above expression in Eq.~[\ref{BKEacousticB1}] and neglect terms of the order: $1\gg q^{2}/\kappa^{2}\gg c_{s} q^{2}/v\kappa^{2}$. We then proceed to find the electric field by applying the normalization condition: $\sum_{\alpha}\langle\chi_{1}^{(\alpha)}\rangle_{0} = 0$, such that the longitudinal electric field produced by the acoustic phonons is given as,

\begin{equation}\label{electricfieldacousticB1}
eE_{1,\parallel} \approx \bigg (\frac{\Gamma_{A}}{\Gamma_{E}} - \frac{i\omega \Gamma_{A}}{\Gamma_{E}^{2}}\bigg)\bigg[\frac{\sum_{\alpha}\bigg(\lambda_{ij}^{(\alpha)} - \frac{\langle \lambda_{ij}\rangle_{0}}{\nu_{0}}\bigg)\langle I^{(\alpha)}\rangle_{0}}{\sum_{\alpha}\langle {\textbf{v}}^{(\alpha)}\cdot \hat{\textbf{q}} R^{(\alpha)}\rangle_{0}}\bigg]  \omega q_{j} u_{i}
\end{equation}

with the corresponding chiral charge imbalance:

\begin{equation}\label{avchiacousticB}
\begin{split}
    \langle\chi_{1}^{(\alpha)}\rangle_{0} &\approx \bigg(\frac{\Gamma_{A}}{\Gamma_{E}}\bigg)^{2}\bigg(1 + \frac{2i\omega}{\Gamma_{E}}\bigg)\bigg[\langle I^{(\alpha)}\rangle_{0} \bigg(\lambda_{ij}^{(\alpha)} - \frac{\langle \lambda_{ij}\rangle_{0}}{\nu_{0}}\bigg)\\ - &\frac{\langle {\textbf{v}}^{(\alpha)}\cdot \hat{\textbf{q}}R^{(\alpha)}\rangle_{0}\sum_{\alpha}\bigg(\lambda_{ij}^{(\alpha)} - \frac{\langle \lambda_{ij}\rangle_{0}}{\nu_{0}}\bigg)\langle I^{(\alpha)}\rangle_{0}}{\sum_{\alpha}\langle {\textbf{v}}^{(\alpha)}\cdot \hat{\textbf{q}}R^{(\alpha)}\rangle_{0}} \bigg]\omega q_{j}u_{i}
\end{split}
\end{equation}

where,
\begin{equation}\label{Int}
\begin{split}
I^{\alpha} &= \bigg[i\textbf{q}\cdot\bigg(\frac{e}{\hbar}(\boldsymbol{\Omega}^{(\alpha)}\cdot\textbf{v}^{(\alpha)})\textbf{B} -\partial_{\textbf{p}}(\textbf{m}^{(\alpha)}\cdot\textbf{B}) \bigg)(R^{(\alpha)})^{2}\\
&\quad + \frac{e}{\hbar}(\textbf{B}\cdot\boldsymbol{\Omega}^{(\alpha)})(R^{(\alpha)})^{2} (\Gamma_{A} - i\omega) -\frac{e}{\hbar}(\textbf{B}\cdot\boldsymbol{\Omega}^{(\alpha)})R^{(\alpha)}\bigg].\\
\end{split}
\end{equation}

In the subsection [\ref{subsec:drag_force_b_acoustic}], we will use expressions of $\langle\chi_{1}^{(\alpha)}\rangle_{0}$ to derive the longitudinal mode for the acoustic phonon. In the next subsection, we lay out the solution of BKE for electron-optical phonon interaction.

\subsubsection{Electron-Optical Phonon Interaction}
\label{subsec:electron-optical_bke_b}

Adapting a similar procedure as in Sec.~[\ref{subsec:acoustic_bke_b}], we will derive in this subsection, expressions for the solution of the BKE in presence of magnetic field for the case of non-polar optical phonons. We begin with Eq.~[\ref{BKEoptical}], apply a perturbative expansion in ${\textbf{E}}$, $\chi^{(\alpha)}$ and $\delta\mu$ till linear order and subsequently, use the expressions for $\delta\mu_{0}$, $\chi_{0}^{(\alpha)}$ from the previous Subsection~\ref{subsec:electron-optical_bke0}, which leads to

\begin{equation}\label{opticalBKEB}
\begin{split}
\chi_{1}^{(\alpha)} &= -I^{(\alpha)}\bigg[i\omega \bigg(g^{(\alpha)} - \frac{\langle g \rangle_{0}}{\nu_{0}}\bigg)\xi_{{\textbf{q}}} + (\Gamma_{A} -\Gamma_{E})\frac{\langle\chi_{0}^{(\alpha)}\rangle_{0}}{\nu_{0}^{(\alpha)}}\bigg]\\
&\quad -e{\textbf{E}}_{1}\cdot \textbf{v}^{(\alpha)} R^{(\alpha)} +  (\Gamma_{A}-\Gamma_E)\frac{\langle\chi_{1}^{(\alpha)}\rangle_0}{\nu^{(\alpha)}_0}R^{(\alpha)}.
\end{split}
\end{equation}
where, we have adhered to the limit of $\kappa^{2}\rightarrow \infty$ and $I^{(\alpha)}$ is given by Eq.~[\ref{Int}]. Next, we take Fermi surface average on both sides and apply the normalization condition $\sum_{\alpha}\langle\chi_{1}^{(\alpha)}\rangle_{0} = 0$, to find an expression for the electric field exerted by the optical phonons in presence of magnetic field,
\begin{equation}\label{E1optical}
eE_{1,\parallel} \approx -i\omega \bigg (1 + \frac{i\Gamma_{A}}{\omega} \bigg)\bigg[\frac{\sum_{\alpha}\bigg(g^{(\alpha)} - \frac{\langle g\rangle_{0}}{\nu_{0}}\bigg)\langle I^{(\alpha)}\rangle_{0}}{\sum_{\alpha}\langle {\textbf{v}}^{(\alpha)}\cdot \hat{\textbf{q}} R^{(\alpha)}\rangle_{0}}\bigg]  \xi_{{\textbf{q}}}
\end{equation}
where, we have used the expression of $\langle\chi_{0}^{(\alpha)}\rangle_{0}$ from Eq.~[\ref{avchi0optical}]. Subsequently, using the above, we find the chiral charge imbalance,
\begin{equation}\label{avchiopticalB1}
\begin{split}
    \langle\chi_{1}^{(\alpha)}\rangle_{0} &\approx -i\omega \bigg(1 + \frac{2i\Gamma_{A}}{\omega}\bigg)\bigg[\langle I^{(\alpha)}\rangle_{0} \bigg(g^{(\alpha)} - \frac{\langle g\rangle_{0}}{\nu_{0}}\bigg)\\
    & - \frac{\langle {\textbf{v}}^{(\alpha)}\cdot \hat{\textbf{q}}R^{(\alpha)}\rangle_{0}\sum_{\alpha}\bigg(g^{(\alpha)} - \frac{\langle g\rangle_{0}}{\nu_{0}}\bigg)\langle I^{(\alpha)}\rangle_{0}}{\sum_{\alpha}\langle {\textbf{v}}^{(\alpha)}\cdot \hat{\textbf{q}}R^{(\alpha)}\rangle_{0}} \bigg]\xi_{{\textbf{q}}}\\
\end{split}
\end{equation}
In the Sec.~[\ref{subsec:drag_force_b_optical}], we will use the expressions for the chiral charge imbalance $\langle\chi_{1}^{(\alpha)}\rangle_{0}$ generated by the longitudinal optical phonons to calculate the drag force exerted by the conduction electrons. Next, we derive the magnetic field dependence in the acoustic dispersion relations.

\subsection{Magnetic field-induced drag force and elasticity equations: Nonreciprocity in the acoustic phonon dispersion}
\label{subsec:drag_force_b_acoustic}

Proceeding in a similar manner as shown in Sec.~[\ref{subsec:drag_force_b0_acoustic}], we plug the expression for $\langle\chi_{1}^{(\alpha)}\rangle_{0}$ given by Eq.~[\ref{avchiacousticB}] in Eq.~[\ref{dragforceacoustic}] and invoke the limit $\kappa^{2}\rightarrow\infty$, to find the drag force in the presence of magnetic field as

\begin{equation}\label{dragacousticB}
\begin{split}
F_{z}^{(ac)}(B) &= i\omega q_{z}^{2}u_{z}\langle I^{(+)} \rangle_{0}(\lambda_{1}^{(+)} - \lambda_{2}^{(-)})^{2} \bigg(\frac{\Gamma_{A}}{\Gamma_{E}}\bigg)^{2} \\
&\times \bigg(1 + \frac{2i\omega}{\Gamma_{E}}\bigg)\bigg[\frac{(\nu_{0}^{(+)})^{2} D^{(+)} - (\nu_{0}^{(-)})^{2} D^{(-)} }{(\nu_{0}^{(+)} + \nu_{0}^{(-)}) (\nu_{0}^{(+)} D^{(+)} + \nu_{0}^{(-)} D^{(-)})}\bigg]
\end{split}
\end{equation} 

Plugging the above in Eq.~[\ref{eq:elasticityacoustic}], we find an expression for the elasticity equation given as,

\begin{equation}
\label{elasticityacousticB}
\begin{split}
   \rho\omega^{2}u_{z} &= q_{z}^{2}s_{33} u_{z} + i\omega q_{z}^{2}u_{z}\langle I^{(+)} \rangle_{0}(\lambda_{1}^{(+)} - \lambda_{2}^{(-)})^{2}\bigg(\frac{\Gamma_{A}}{\Gamma_{E}}\bigg)^{2}\\ &\bigg(1 + \frac{2i\omega}{\Gamma_{E}}\bigg)\bigg[\frac{(\nu_{0}^{(+)})^{2} D^{(+)} - (\nu_{0}^{(-)})^{2} D^{(-)} }{(\nu_{0}^{(+)} + \nu_{0}^{(-)}) (\nu_{0}^{(+)} D^{(+)} + \nu_{0}^{(-)} D^{(-)})}\bigg]
\end{split}
\end{equation}

From Eq.~(\ref{Int}), we derive an expression $\langle I^{(+)}\rangle_{0}$ using integrals given in Appendix (\ref{sec:int_acoustic}) such that
\begin{equation}\label{Iacoustic}
\begin{split}
\langle I^{(+)} \rangle_{0} 
&\simeq \frac{{\textbf{q}}\cdot{\textbf{B}}}{2\pi^{2}\hbar^{2}}\frac{e|C|}{\Gamma_{A}^{2}}\bigg[\frac{i}{2}\bigg(1-\frac{2\omega^{2}}{\Gamma_{A}^{2}}\bigg) - \frac{\omega}{\Gamma_{A}}\bigg]\\
&\approx\quad \frac{{i\textbf{q}}\cdot{\textbf{B}}}{4\pi^{2}\hbar^{2}}\frac{e|C|}{\Gamma_{A}^{2}}\bigg[1+ \frac{2i\omega}{\Gamma_{A}}\bigg].
\end{split}
\end{equation}

Using the above expression in Eq.~(\ref{elasticityacousticB}), we find the magnetic field correction to the longitudinal acoustic phonon dispersion relation,

\begin{equation}\label{dispersionacousticB}
\begin{split}
   0 &= q_{z}^{2}s_{33} - \rho\omega^{2} - \frac{\omega q_{z}^{2} |q_{z}| eB_{z}|C|}{4\pi^{2}\hbar^{2}\Gamma_{E}^{2}} \bigg(\lambda_{1}^{(+)} - \lambda_{1}^{(-)}\bigg)^{2}\\
   &\bigg(\frac{(\nu_{0}^{(+)})^{2} D^{(+)} - (\nu_{0}^{(-)})^{2} D^{(-)} }{(\nu_{0}^{(+)} + \nu_{0}^{(-)}) (\nu_{0}^{(+)} D^{(+)} + \nu_{0}^{(-)} D^{(-)})}\bigg) \bigg(1 + \frac{2i\omega}{\Gamma_{E}}\bigg)
\end{split}
\end{equation}

The magnetic-field corrections to the acoustic phonon dispersion are odd in $q_z$ as well as in $B_z$, and proportional to $|C|$. This confirms the existence of a phonon magnetochiral effect of purely band-geometric origin. 

\begin{figure*}[t]
    \centering
    \includegraphics[width=0.98\textwidth]{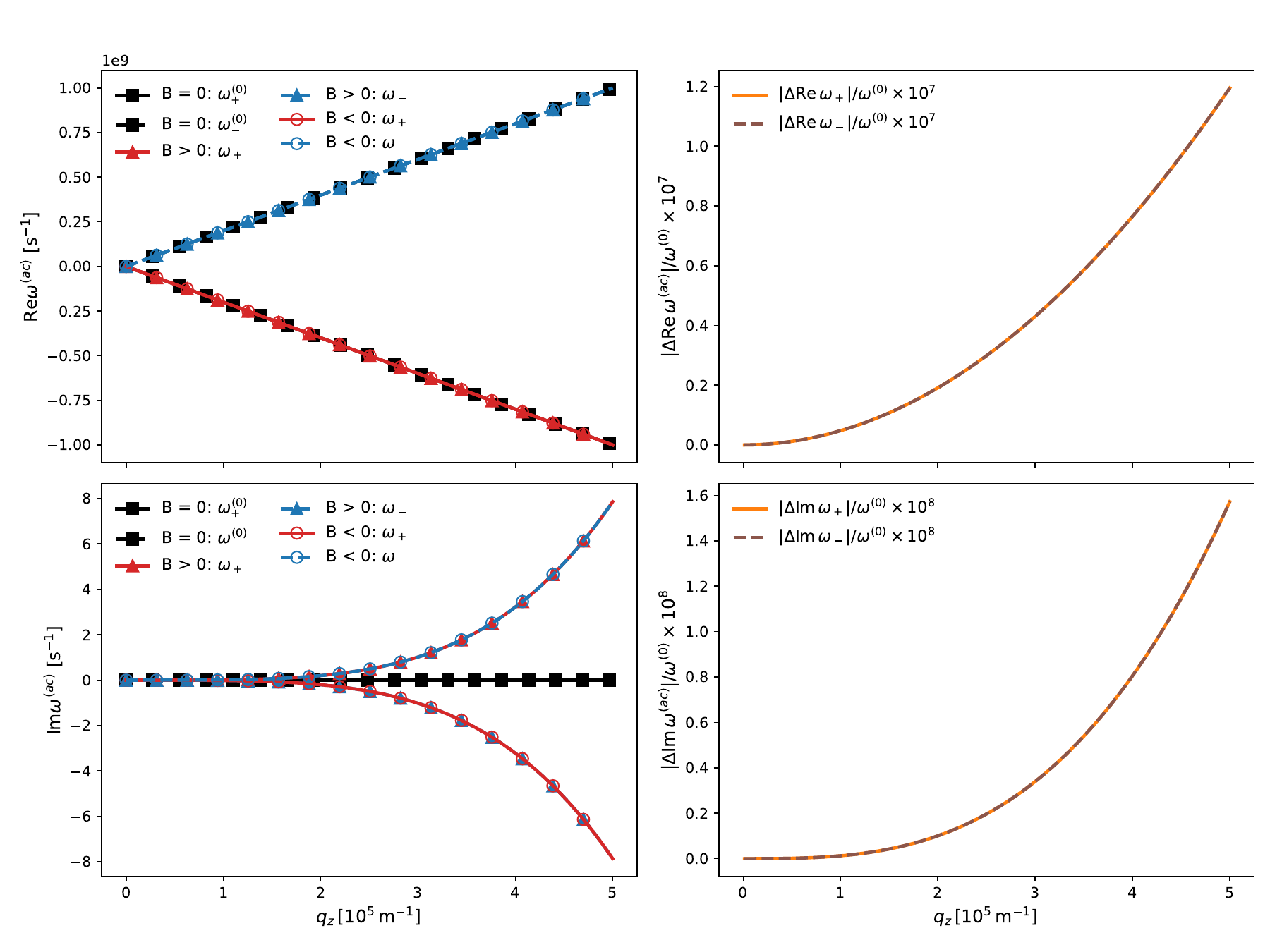}
    \caption{Acoustic phonon dispersion in a chiral Weyl semimetal for zero magnetic field ($\omega^{(0)}$) and for finite fields $B>0$ and $B<0$, the latter corresponding to phonon propagation parallel and antiparallel to $\mathbf{B}$. The top-left panel displays the real part of the dispersion, where the zero-field curve $\omega^{(0)}$ represents the electron–phonon renormalized longitudinal mode. A finite magnetic field introduces a small magnetochiral splitting between the $\omega_{\pm}$ (for $B>0$) and $\omega_{\pm}$ (for $B<0$) branches. The magnitude of this shift, shown in the top-right panel, is small but experimentally resolvable \cite{senguptapmce}. In contrast, the imaginary part of the dispersion (bottom-left) exhibits a substantially larger nonreciprocal correction, as quantified in the bottom-right panel. 
}

    \label{fig:real_acoustic}
\end{figure*}
In Fig.~\ref{fig:real_acoustic}, we show the numerical results for the acoustic dispersion relation along with the variation of the magnetochiral factor:

\begin{equation}
\label{eq:magnetochiral}
\frac{|\Delta\omega|}{\omega^{(0)}} = \frac{|\omega(B>0) - \omega(B<0)|}{\omega(B=0)}.
\end{equation}
Here, we use the following parameters\cite{senguptapmce}: for a magnetic field $B=1$T, $\nu_{0}^{(\alpha)} = (\epsilon_{F}^{(\alpha)})^{2}/(v_{F}^{(\alpha)})^{3}$, with $\epsilon_{F}^{(+)} = 20$meV,  $\epsilon_{F}^{(-)} = 5$meV, $v_{F}^{(+)} = 10^{5}$m/s, $v_{F}^{(-)} = 1.5\times10^{5}$m/s, $\lambda_{1}^{(+)} = 1.75$eV, $\lambda_{1}^{(-)} = 1.25$eV, $\Gamma_{E} = 10^{-2}$meV, $\rho=10^4\,{\rm kg/m}^3$ and $q_{z} < 5 \times 10^{5}$ m$^{-1}$.

Figure~\ref{fig:real_acoustic} shows the full acoustic phonon dispersion at zero magnetic field together with the two field-reversed branches obtained for $B>0$ and $B<0$. The zero-field curve $\omega^{(0)}$ reflects the electron–phonon renormalized longitudinal mode obtained in Subsection~\ref{subsec:drag_force_b0_acoustic}. Upon applying a magnetic field, the real part of the dispersion acquires a small but finite magnetochiral shift, producing a splitting between the $\omega_{\pm}$ (for $\mathbf{q}\parallel\mathbf{B}$) and $\omega_{\pm}$ (for $\mathbf{q}\!\parallel\!-\,\mathbf{B}$) branches. Although this shift is minute, the right-hand panels make clear that its magnitude grows quadratically with $q_z$ and remains within the resolution of modern ultrasound probes \cite{senguptapmce}. In contrast, the imaginary part of the dispersion exhibits a significantly stronger nonreciprocal response: the attenuation coefficients for $\omega_{\pm}$ and $\omega_{\pm}$ separate by an amount that is one to two orders of magnitude larger than the corresponding shift in the real part. 

Next, we show the magnetic field dependence of the optical phonon dispersion relations.

\subsection{Magnetic field dependence of drag force and elasticity equations: Non-reciprocity in optical phonon dispersion relations}
\label{subsec:drag_force_b_optical}

\begin{figure*}[t]
    \centering
    \includegraphics[width=0.98\textwidth]{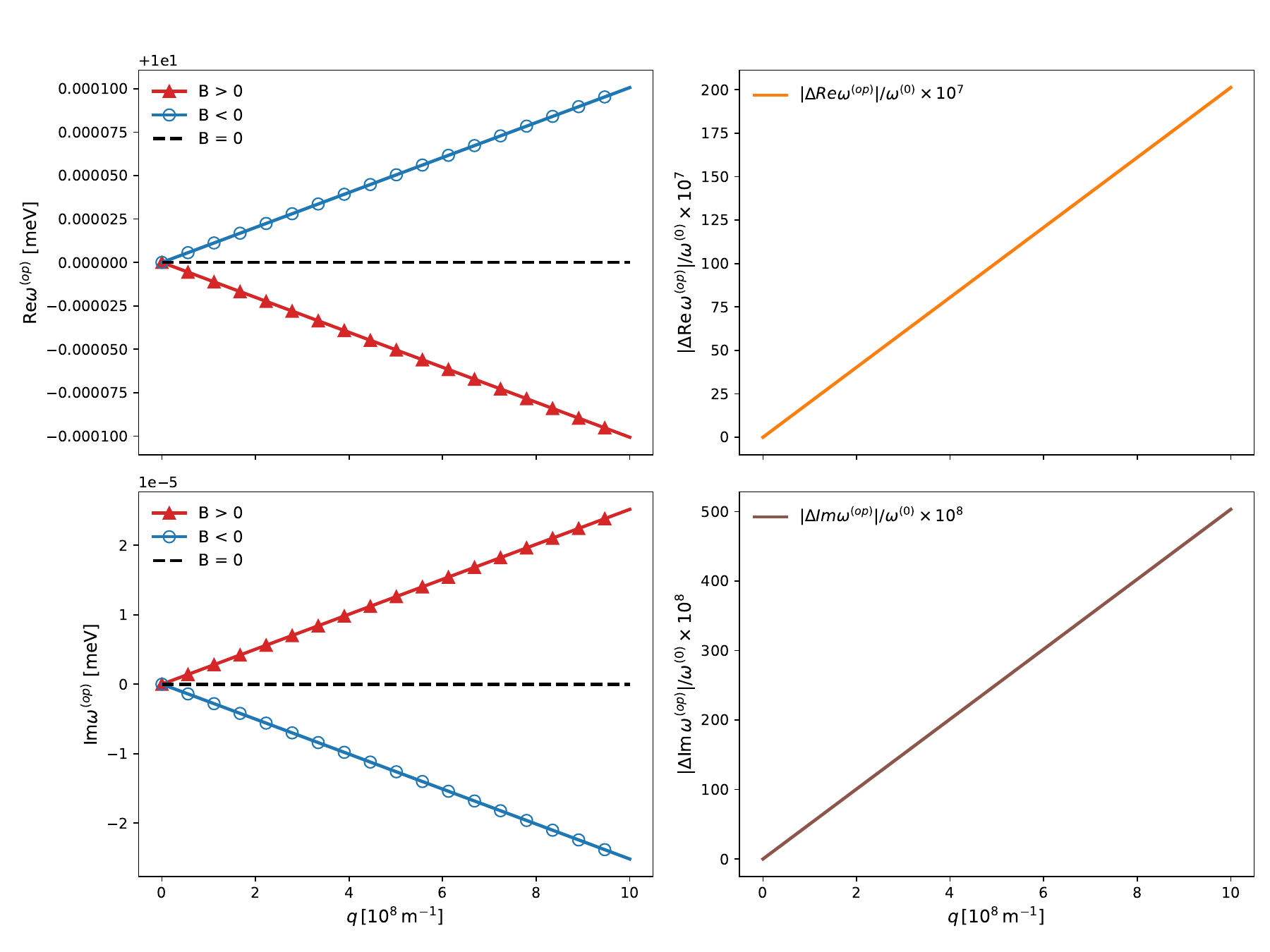}
    \caption{Optical phonon dispersion for zero magnetic field and for finite fields $B>0$ and $B<0$ in a chiral Weyl semimetal. The zero-field dispersion ($\omega^{(0)}$) represents the electron–phonon renormalized non-polar optical mode. Applying a magnetic field produces a linear magnetochiral splitting between the $\omega(B>0)$ and $\omega(B<0)$ branches: the real part (top-left) shifts by a small but finite amount, while the imaginary part (bottom-left) exhibits a significant directional asymmetry in attenuation. The right-hand panels quantify the magnetochiral effects, showing that the optical-phonon PMCE increases linearly with $q$.
}
    \label{fig:real_optical}
\end{figure*}

In the presence of magnetic field, following Eq.~[\ref{dragforceoptic}] and Eq.~[\ref{avchiopticalB1}], we derive,

\begin{equation}\label{FdragopticalB}
\begin{split}
F^{(op)}(B) &\simeq -\frac{\mathcal{V}}{N}(g^{(+)} - g^{(-)})^{2}(i\omega - 2i\Gamma_{A}) \langle I^{(+)}\rangle_{0}\\& \bigg[\frac{(\nu_{0}^{(+)})^{2} D^{(+)} - (\nu_{0}^{(-)})^{2} D^{(-)} }{(\nu_{0}^{(+)} + \nu_{0}^{(-)}) (\nu_{0}^{(+)} D^{(+)} + \nu_{0}^{(-)} D^{(-)})}\bigg]\xi_{{\textbf{q}}}.
\end{split}
\end{equation}

Using the above expression in the elasticity equation, we find an expression for the optical phonon dispersion relation,

\begin{equation}\label{elasticityopticalB}
\begin{split}
(\omega^{2} -\omega_{0}^{2}) \xi_{{\textbf{q}}} &=-\frac{\mathcal{V}}{NM}(g^{(+)} - g^{(-)})^{2} (i\omega - 2i\Gamma_{A}) \langle I^{(+)}\rangle_{0} \\& \bigg[\frac{(\nu_{0}^{(+)})^{2} D^{(+)} - (\nu_{0}^{(-)})^{2} D^{(-)} }{(\nu_{0}^{(+)} + \nu_{0}^{(-)}) (\nu_{0}^{(+)} D^{(+)} + \nu_{0}^{(-)} D^{(-)})}\bigg]\xi_{{\textbf{q}}}.
\end{split}
\end{equation}

Once again, within the long-wavelength approximation, the above equation reduces to,

\begin{equation}\label{elasticityopticalB}
\begin{split}
(\omega^{2} -\omega_{0}^{2}) \xi_{{\textbf{q}}} &=-\frac{\Delta^{2}}{\rho' a^{2}} (i\omega - 2i\Gamma_{A}) \langle I^{(+)}\rangle_{0} \\
&\times \bigg[\frac{(\nu_{0}^{(+)})^{2} D^{(+)} - (\nu_{0}^{(-)})^{2} D^{(-)} }{(\nu_{0}^{(+)} + \nu_{0}^{(-)}) (\nu_{0}^{(+)} D^{(+)} + \nu_{0}^{(-)} D^{(-)})}\bigg]\xi_{{\textbf{q}}}.
\end{split}
\end{equation}

Using the expressions for the integrals given in Appendix~(\ref{sec:int_optical}), we find
\begin{equation}\label{opticalI}
\langle I^{(+)} \rangle_{0} \approx -\frac{ieBq|C|}{4\pi^{2}\hbar^{2}\omega^{2}} \bigg(1 - \frac{2i\Gamma_{A}}{\omega}\bigg).
\end{equation}

Using Eq.~(\ref{opticalI}) in Eq.~(\ref{elasticityopticalB}), we find the final expression of the optical phonon dispersion relation in presence of magnetic field,

\begin{equation}
\begin{split}
(\omega^{2} -\omega_{0}^{2})\xi_{{\textbf{q}}} &=  \frac{\Delta^{2}}{\rho' a^{2}} \frac{eBq|C|}{4\pi^{2}\hbar^{2}} \bigg[\bigg(\frac{2\Gamma_A}{\omega^{2}} -\frac{1}{\omega}\bigg) + \frac{2i\Gamma_A}{\omega^{2}}\bigg] \\
&\times \bigg[\frac{(\nu_{0}^{(+)})^{2} D^{(+)} - (\nu_{0}^{(-)})^{2} D^{(-)} }{(\nu_{0}^{(+)} + \nu_{0}^{(-)}) (\nu_{0}^{(+)} D^{(+)} + \nu_{0}^{(-)} D^{(-)})}\bigg]\xi_{{\textbf{q}}},
\end{split}
\end{equation}

The above equation is numerically solved to obtain the optical phonon dispersion relations. We have used parameters similar to the ones for acoustic phonons, except the bare optical phonon energy $\hbar\omega_{0} = 10 meV$, the lattice spacing is $a = 4\AA$ and $\Delta = 5$eV. In Fig.~[\ref{fig:real_optical}], we display both the real and imaginary part of the optical phonon dispersion along with the magnetochiral effect following Eq.~[\ref{eq:magnetochiral}]. In contrast to the acoustic case, the optical branch exhibits a strictly linear in $q$ splitting between the $\omega(B>0)$ and $\omega(B<0)$ modes once a magnetic field is applied. The real and imaginary part of the dispersion displays a finite magnetochiral effect.

\section{Discussion \& Outlook}
\label{sec:discussion}

In this work, we have developed a comprehensive theoretical framework for the phonon magnetochiral effect (PMCE) arising from band-geometric properties of chiral Weyl fermions. Using a semiclassical kinetic approach, we incorporated the full non-equilibrium electron–phonon coupling within the Boltzmann kinetic equation and derived the resulting phonon dispersion relations through the elasticity equations. This methodology enabled us to treat acoustic and non-polar optical phonons on equal footing and to obtain closed-form analytic expressions for their nonreciprocal dispersions.

Our results reveal that inequivalent Weyl nodes differing in Fermi energies, Fermi velocities, Berry curvature, and orbital magnetic moment—produce distinct corrections to the real and imaginary parts of the phonon spectrum. These node-dependent effects manifest differently in acoustic and optical branches: the acoustic mode exhibits a subtle magnetochiral frequency shift but a strongly enhanced nonreciprocal attenuation, whereas the optical mode displays a linear in $q$ magnetochiral splitting. Both behaviors reflect the underlying dynamical anomaly generated by phonons propagating along a magnetic field, which induces a chiral population imbalance that feeds back on lattice dynamics. The PMCE thus emerges as a sensitive, non-electronic probe of topological band geometry and anomaly-related physics.

While the present study focuses on \emph{non-polar} optical phonons whose vanishing Born effective charge ensures the absence of LO--TO splitting, an important next step is to extend the formalism to \emph{polar} optical modes\cite{rinkel2017, rinkel2019}. In polar crystals, long-range Coulomb interactions generate macroscopic electric fields and modify the phonon spectrum through LO--TO splitting. Incorporating these electrodynamic effects into our semiclassical framework would allow for a unified description of magnetochiral phenomena in both polar and non-polar materials, including the interplay among lattice polarization, Berry curvature, and Weyl-node asymmetry. Such generalizations lie beyond the present scope but represent compelling directions for future work, particularly in systems where strong electron–phonon coupling and lattice polarization coexist with topological electronic structure.

Another interesting extension of our work could be foreseen in the field of straintronics and non-Reciprocal transport in heat and sound. Strain engineering in quantum materials has the potential to revolutionize heat dissipation in next-generation electronics and energy-efficient devices. Building on the theoretical formalism described in this work, one could explore strain engineering \cite{cortijo, pikulin, sumiyoshi, yago} to induce non-reciprocal transport phenomena in heat and sound. Applications of such phenomena would include directional heat transport, acoustic waveguides, and straintronic devices \cite{nomura, toposound, ding}. A key challenge lies in optimizing strain-induced non-reciprocal transport to achieve scalable heat management systems.

\section*{Acknowledgments}
This project started at the Institut Quantique, Universit$\acute{e}$ de Sherbrooke with support from Merit Scholarship, Fonds qu$\acute{e}$b$\acute{e}$cois de la recherche sur la nature et les technologies (FRQNT), and thereafter SS has subsequently received financial support from Brandeis University. 

\appendix
\section{Expressions for integrals derived for acoustic PMCE}
\label{sec:int_acoustic}

In this section, we provide the expressions for the necessary integrals that are used in the main text involving the acoustic phonons (please see Sec.~[\ref{subsec:drag_force_b_acoustic}]). In the following, we have used Eqs.~[\ref{v0}, \ref{Om} \ref{m}, \ref{gradm}, \ref{Racoustic0}] such that,
\begin{equation}\label{intvalues}
\begin{split}
    -\frac{e}{\hbar}\langle(\textbf{B}\cdot\boldsymbol{\Omega}^{(\alpha)})R^{(\alpha)}\rangle_{0} & \approx \frac{eBq|C|\alpha}{6\pi^{2}\hbar^{2}} \frac{1}{\Gamma_{A}^{3}}\bigg(\frac{i\Gamma_{A}}{2} - \omega\bigg) \\
    &\quad\approx \frac{ieBq|C|\alpha}{12\pi^{2}\hbar^{2}} \frac{1}{\Gamma_{A}^{2}}\bigg(1+\frac{2i\omega}{\Gamma_{A}}\bigg)\\
\end{split}
\end{equation}
\begin{equation}
\begin{split}
    \frac{e}{\hbar}\langle(\textbf{B}\cdot\boldsymbol{\Omega}^{(\alpha)})(R^{(\alpha)})^{2}\rangle_{0} & \approx \frac{eBq|C|\alpha}{2\pi^{2}\hbar^{2}}\frac{1}{\Gamma_{A}^{3}} \bigg(-\frac{i}{3} +\frac{\omega}{\Gamma_{A}}\bigg)\\
    &\quad\approx -\frac{ieBq|C|\alpha}{6\pi^{2}\hbar^{2}}\frac{1}{\Gamma_{A}^{3}} \bigg(1 +\frac{3i\omega}{\Gamma_{A}}\bigg)\\
\end{split}
\end{equation}
\begin{equation}
\begin{split}
    i\textbf{q}\cdot\frac{e}{\hbar}\langle(\boldsymbol{\Omega}^{(\alpha)}\cdot\textbf{v}^{(\alpha)})\textbf{B} (R^{(\alpha)})^{2}\rangle_{0} & \approx \frac{eBq|C|\alpha}{2\pi^{2}\hbar^{2}\Gamma_{A}^{2}}\bigg(\frac{i}{2} -\frac{\omega}{\Gamma_{A}}\bigg) \\
    &\approx \frac{ieBq|C|\alpha}{4\pi^{2}\hbar^{2}\Gamma_{A}^{2}}\bigg(1 +\frac{2i\omega}{\Gamma_{A}}\bigg) \\
\end{split}
\end{equation}
\begin{equation}
\begin{split}
    -i\textbf{q}\cdot \langle\partial_{\textbf{p}}(\textbf{m}^{(\alpha)}\cdot\textbf{B}) (R^{(\alpha)})^{2}\rangle_{0} &\approx \frac{eBq|C|\alpha}{6\pi^{2}\hbar^{2}\Gamma_{A}^{2}}\bigg(\frac{i}{2} -\frac{\omega}{\Gamma_{A}}\bigg)\\
    &\approx \frac{ieBq|C|\alpha}{12\pi^{2}\hbar^{2}\Gamma_{A}^{2}}\bigg(1 +\frac{2i\omega}{\Gamma_{A}}\bigg)\\
\end{split}
\end{equation}

\section{Expressions for integrals derived for optical PMCE}
\label{sec:int_optical}
Similarly, for the optical phonons, using Eq.~[\ref{v0}, \ref{Om} \ref{m}, \ref{gradm}, \ref{Roptic0}], the expressions for the integrals used in Sec.~[\ref{subsec:drag_force_b_optical}] are found as,
\begin{equation}\label{intvaluesoptic}
\begin{split}
    -\frac{e}{\hbar}\langle(\textbf{B}\cdot\boldsymbol{\Omega}^{(\alpha)})R^{(\alpha)}\rangle_{0} & \approx -\frac{ieBq|C|\alpha}{12\pi^{2}\hbar^{2}} \frac{1}{\omega^{2}}\bigg(1-\frac{2i\Gamma_{A}}{\omega} \bigg) \\
\end{split}
\end{equation}

\begin{equation}
\begin{split}
    \frac{e}{\hbar}\langle(\textbf{B}\cdot\boldsymbol{\Omega}^{(\alpha)})(R^{(\alpha)})^{2}\rangle_{0} & \approx -\frac{eBq|C|\alpha}{6\pi^{2}\hbar^{2}}\frac{1}{\omega^{3}} \bigg(1-\frac{3i\Gamma_{A}}{\omega}\bigg)\\
\end{split}
\end{equation}

\begin{equation}
\begin{split}
 i\textbf{q}\cdot\frac{e}{\hbar}\langle(\boldsymbol{\Omega}^{(\alpha)}\cdot\textbf{v}^{(\alpha)})\textbf{B} (R^{(\alpha)})^{2}\rangle_{0} & \approx -\frac{ieBq|C|\alpha}{4\pi^{2}\hbar^{2}\omega^{2}}\bigg(1 - \frac{2i\Gamma_{A}}{\omega} \bigg)\\ 
\end{split}
\end{equation}

\begin{equation}
\begin{split}
    -i\textbf{q}\cdot \langle\partial_{\textbf{p}}(\textbf{m}^{(\alpha)}\cdot\textbf{B}) (R^{(\alpha)})^{2}\rangle_{0} &\approx -\frac{ieBq|C|\alpha}{12\pi^{2}\hbar^{2}\omega^{2}}\bigg(1 - \frac{2i\Gamma_{A}}{\omega} \bigg)\\
\end{split}
\end{equation}


\clearpage
\bibliography{refs.bib}
\end{document}